\documentclass[aps,prb,amsmath,amssymb,reprint,superscriptaddress]{revtex4-2}

\usepackage{graphicx}
\usepackage{amsmath,amssymb}
\usepackage{xcolor}
\usepackage{color,soul}

\usepackage{geometry}
\usepackage{hyperref}
\usepackage{listings}
\usepackage{xcolor}
\usepackage{url}
\usepackage{caption}
\usepackage{float}
\usepackage{listings}
\usepackage{xcolor}

\begin{document}

\title{Size-Dependent Structural Motifs in Ag$_n$Mo (n = 2–13) Clusters: From Planar to Icosahedral Architectures}

\author{Samantha Ortega-Flores}
\affiliation{Centro de Investigaci\'on en Ciencias Aplicadas y Tecnolog\'ia Avanzada, Unidad Legaria, Instituto Polit\'ecnico Nacional, M\'exico}
\author{P. L. Rodr\'iguez-Kessler}
\email{plkessler@cio.mx}
\affiliation{Centro de Investigaciones en \'Optica A.C., Loma del Bosque 115, Lomas del Campestre, Leon, 37150, Guanajuato, Mexico}

\date{\today}

\begin{abstract}
We present a comprehensive density functional theory (DFT) study of Mo-doped silver clusters Ag$_n$Mo ($n=1$–14), focusing on their structural, electronic, and bonding properties. Global optimization reveals an evolution from planar and low-symmetry isomers in small clusters to compact three-dimensional geometries with higher symmetry, culminating in a highly stable icosahedral structure at $n=12$. Binding energy and second-order energy difference analyses identify $n=12$ as a “magic number” cluster exhibiting enhanced thermodynamic stability and a pronounced HOMO–LUMO gap, indicative of electronic shell closure. Bond length analysis shows relatively constant Ag–Mo distances alongside a size-dependent increase in Ag–Ag bond lengths, reflecting the growth of metallic bonding networks. Hirshfeld charge analysis reveals significant charge transfer from Ag to Mo in small clusters, which decreases with size as the system transitions toward delocalized metallic bonding. These findings provide detailed insights into the size-dependent interplay of geometry, bonding, and electronic structure in Ag$_n$Mo clusters, with implications for their catalytic and material applications.
\end{abstract}


\maketitle

\section{Introduction}

The structural and electronic properties of atomically precise metal clusters have attracted significant attention due to their size-dependent behavior and potential applications in catalysis, optics, and nanomaterials.\cite{10.1063/5.0204606} Among such systems, doped noble metal clusters represent a particularly rich area of study, since transition metal dopants can dramatically alter the geometric, bonding, and electronic characteristics of the host cluster.\cite{D1NA00077B} Silver clusters (Ag$_n$) are especially interesting in this regard due to their optical properties, which make them suitable for both fundamental investigations and technological applications.\cite{doi:10.1021/acs.accounts.7b00295,GRIGORYAN2013197}

A significant amount of research in the field of gas-phase clusters has been devoted to the influence of doping on the properties of clusters, such as their structures, optical responses,\cite{Chaudhary2024,Li2022_AgCuClusters} and reactivity.\cite{RODRIGUEZKESSLER2024141588,doi:10.1021/acs.jpcc.7b05048} These studies have shown that a single dopant atom can strongly influence the properties of clusters. Moreover, studies have shown that dopants can induce magic number behavior in host clusters, where certain sizes exhibit enhanced stability due to geometric shell closure, electronic shell filling, or both.\cite{RODRIGUEZKESSLER201555,Nguyen2023_AgCoClusters,RODRIGUEZKESSLER2025122349,Tian2025_PdAgClusters} Identifying such magic number clusters and characterizing their bonding, charge distribution, and density of states provides valuable insight into the underlying principles governing cluster stability and reactivity.\cite{doi:10.1021/acs.jpca.6b00224,RODRIGUEZKESSLER2019141,MORATOMARQUEZ2020137677,D1CP05410D,RODRIGUEZKESSLER2023121620,10.1002/cphc.202401118}

For noble metal clusters, it was found that molybdenum and niobium doping reduced the CO adsorption energy, which is important to increase the CO tolerance in fuel cell applications.\cite{KAYDASHEV2015133} Doping silver clusters with early transition metals introduces localized $d$-state interactions and modifies the charge distribution, with implications in the stability and reactivity.\cite{Xiong2017,https://doi.org/10.1002/jcc.27197} The incorporation of a single Mo atom into Ag$_n$ clusters can lead to a wide variety of structural motifs, ranging from planar to compact three-dimensional geometries, depending on the size of the silver host and the nature of the Ag–Mo bonding. Understanding how Mo doping influences the evolution of cluster geometry and electronic structure is essential for the rational design of cluster-based catalysts and materials with tailored properties.

In this work, we perform a comprehensive density functional theory (DFT) study of Mo-doped silver clusters Ag$_n$Mo ($n = 1$–14). Using global optimization techniques, we identify the most stable isomers and analyze their structural evolution, electronic properties, and bonding characteristics. We find that the Ag$_{12}$Mo cluster stands out as a particularly stable species, adopting a high-symmetry icosahedral geometry and exhibiting clear signatures of electronic shell closure. Detailed analysis of binding energies, HOMO--LUMO gaps, bond metrics, charge transfer, and density of states reveals how the interplay of atomic size, electronic effects, and coordination environment determines the properties of these clusters. Our findings provide fundamental insights into the structural motifs and stability trends of Mo-doped silver clusters, with implications for their future use in nanoscale materials design.

\section{Computational Details}


This study employs a standard Basin Hopping (BH) global optimization algorithm implemented entirely in Python, interfaced with first-principles calculations via ORCA software. The BH scheme explores the potential energy surface by generating random perturbations to atomic coordinates and performing full structural relaxations using density functional theory (DFT). Energy evaluation for each perturbed configuration is carried out using a PBE0/LANL2DZ level of theory.

Acceptance of new configurations is determined using the Metropolis criterion: if the new energy is lower than the previous one, the configuration is always accepted; otherwise, it is accepted with a probability proportional to $exp(-\Delta{E}/kT)$, where $\Delta{E}$ is the energy difference and T is the simulation temperature. The simulation is designed to maintain an acceptance rate close to 50\% (0.5), ensuring an effective balance between exploration and convergence. All calculations are executed through SLURM, with job scripts files automatically generated and managed by the Python workflow.

The most stable structures identified through the Basin Hopping search are subsequently re-optimized using a higher level of theory to ensure more accurate energetics and geometries. These final optimizations are performed with Orca 6.0.0 code.\cite{10.1063/5.0004608} using the hybrid PBE0 exchange-correlation functional and the Def2-TZVP basis set,\cite{10.1063/1.478522,B508541A} considering multiple spin states to account for potential spin crossover and electronic effects. Atomic positions are self-consistently relaxed through a Quasi-Newton method employing the BFGS algorithm. The SCF convergence criteria for geometry optimizations are achieved when the total energy difference is smaller than 10$^{-8}$ au, by using the TightSCF keyword in the input. The  Van  der  Waals  interactions  are  included in the exchange-correlation functionals with empirical dispersion corrections of Grimme DFT-D3(BJ). 

The following expressions were employed to analyze the energetic and electronic properties of Ag$_n$Mo clusters. The binding energy per atom was calculated using:
\begin{equation}
    E_b(n) = \frac{n E_{\text{Ag}} + E_{\text{Mo}} - E(\text{Ag}_n\text{Mo})}{n+1},
\end{equation}
where $E_{\text{Ag}}$ and $E_{\text{Mo}}$ are the total energies of isolated Ag and Mo atoms, respectively, and $E(\text{Ag}_n\text{Mo})$ is the total energy of the cluster. 

To evaluate relative stability across sizes, the second-order energy difference was computed as:
\begin{equation}
    \Delta_2 E(n) = E(n+1) + E(n-1) - 2E(n),
\end{equation}
which highlights cluster sizes that are energetically more favorable compared to their neighbors (positive values indicate increased stability).

The HOMO--LUMO gap, a descriptor of electronic stability and reactivity, was obtained from the Kohn--Sham orbital energies as:
\begin{equation}
    E_{\text{gap}} = E_{\text{LUMO}} - E_{\text{HOMO}},
\end{equation}
where $E_{\text{HOMO}}$ and $E_{\text{LUMO}}$ are the energies of the highest occupied and lowest unoccupied molecular orbitals, respectively. A large gap is generally associated with higher chemical stability and lower reactivity.

To quantify the overall bonding environment in the MoAg$_n$ clusters, we computed the average bond distance $d_{\mathrm{av}}$ as a weighted mean of all Ag--Mo and Ag--Ag bond lengths. For each cluster, this value was calculated using the expression:

\begin{equation}
d_{\mathrm{av}} = \frac{n_{\mathrm{Ag{-}Mo}} \cdot d_{\mathrm{Ag{-}Mo}} + n_{\mathrm{Ag{-}Ag}} \cdot d_{\mathrm{Ag{-}Ag}}}{n_{\mathrm{Ag{-}Mo}} + n_{\mathrm{Ag{-}Ag}}}
\end{equation}

\noindent where $d_{\mathrm{Ag{-}Mo}}$ and $d_{\mathrm{Ag{-}Ag}}$ denote the average bond lengths for Ag--Mo and Ag--Ag pairs, respectively, and $n_{\mathrm{Ag{-}Mo}}$, $n_{\mathrm{Ag{-}Ag}}$ are the corresponding total numbers of such bonds in the optimized geometry. This descriptor provides a global measure of the bonding compactness in each cluster and was used to generate the smooth green curve in Figure~\ref{fig2}.

\section{Results}

\begin{figure*}[htbp]
    \centering
    \includegraphics[width=0.9\textwidth]{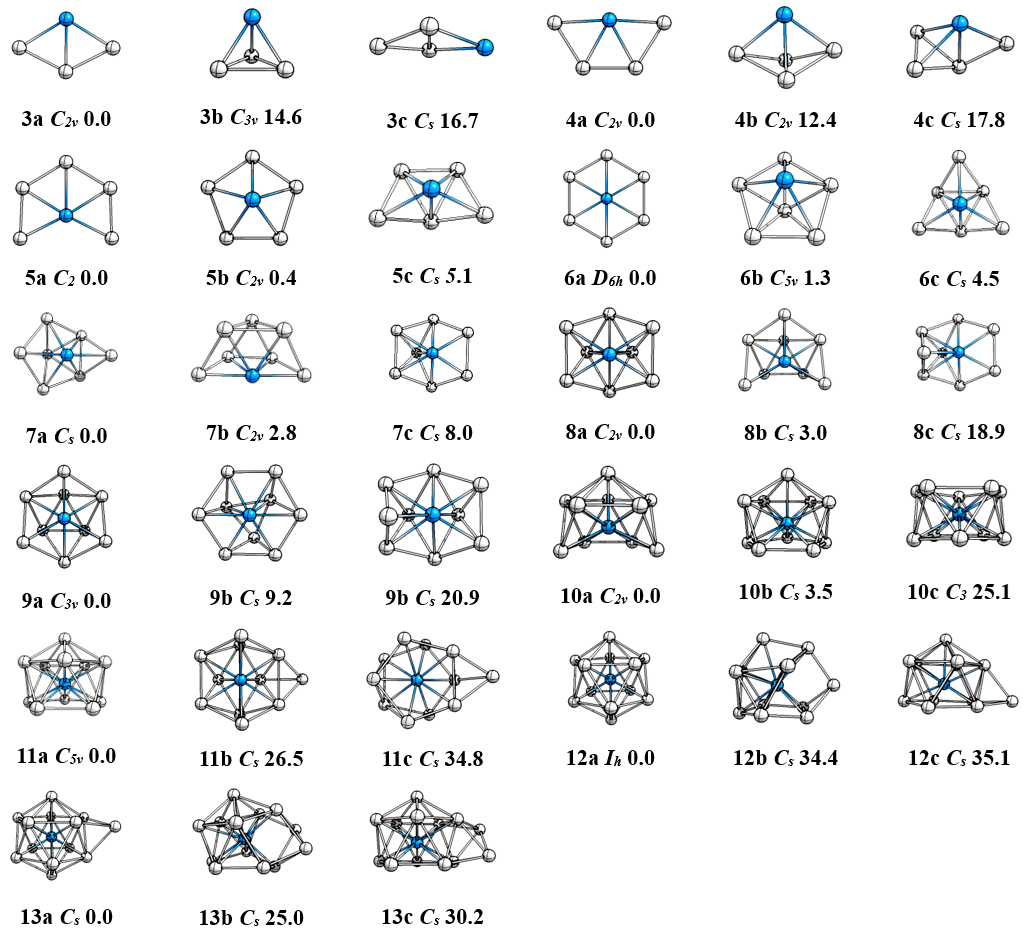}
    \caption{Most stable structures of Ag$_n$Mo (n = 1–13) clusters at the PBE0/Def2-TZVP level. For each cluster, the relative energy (in kcal/mol) and spin multiplicity are given.} 
    \label{fig_struc}
\end{figure*}

Figure~\ref{fig_struc} presents the three lowest-energy isomers obtained from DFT optimization for Ag$_n$Mo clusters with $n$ ranging from 3 to 13 (number of Ag atoms). For each cluster size, the point group symmetry and the relative energy (in kcal/mol) of each isomer are indicated. The global minimum for each size corresponds to the structure with $0.0$ kcal/mol. As the cluster size increases, a clear trend toward more compact and highly coordinated geometries is observed. For small sizes ($n = 3$--$6$), the structures tend to be planar or quasi-planar with low symmetries such as $C_1$, $C_2$, and $C_v$ subgroups, reflecting the geometric constraints imposed by the small number of atoms. For example, the lowest-energy isomer of Ag$_3$Mo adopts a bent $C_{2v}$ configuration, whereas Ag$_4$Mo and Ag$_5$Mo favor slightly more symmetrical structures with increasing coordination around the Mo atom.

From Ag$_6$Mo onwards, the clusters transition to more three-dimensional geometries. Notably, the global minima of Ag$_7$Mo through Ag$_{11}$Mo display symmetries such as $C_s$, $C_2$, and $C_1$, suggesting distorted 3D motifs, possibly due to subtle energetic competitions between strain minimization and electron delocalization around the Mo center. Interestingly, some isomers of intermediate size exhibit relatively high relative energies (e.g., \textbf{12b} with 34.4 kcal/mol), indicating that several local minima are accessible but energetically less favorable. The Ag$_{12}$Mo cluster stands out for displaying an $I_h$ symmetric isomer as its global minimum, indicating a nearly icosahedral structure. This reflects a strong tendency for closed-shell icosahedral motifs at this size, likely stabilized by geometric and electronic shell closure effects. The high symmetry of this structure correlates with increased stability, as evidenced by the relatively large energy gaps between the $I_h$ minimum and the higher-energy $C_1$ and $C_s$ isomers. For Ag$_{13}$Mo, the three isomers show lower symmetries ($C_s$), possibly due to Jahn-Teller distortions or specific Mo--Ag interactions that break the ideal symmetry. Overall, the results highlight a structural evolution from low-symmetry, planar geometries to more symmetric, compact 3D motifs as the number of Ag atoms increases. The emergence of an icosahedral minimum at $n = 12$ suggests a particularly favorable size for shell closure in Ag$_n$Mo clusters. These findings provide valuable insight into the size-dependent stability and structural motifs of Mo-doped silver clusters, which are relevant for understanding their catalytic and electronic properties.


\begin{figure}[h!]
    \centering
    \includegraphics[width=0.47\textwidth]{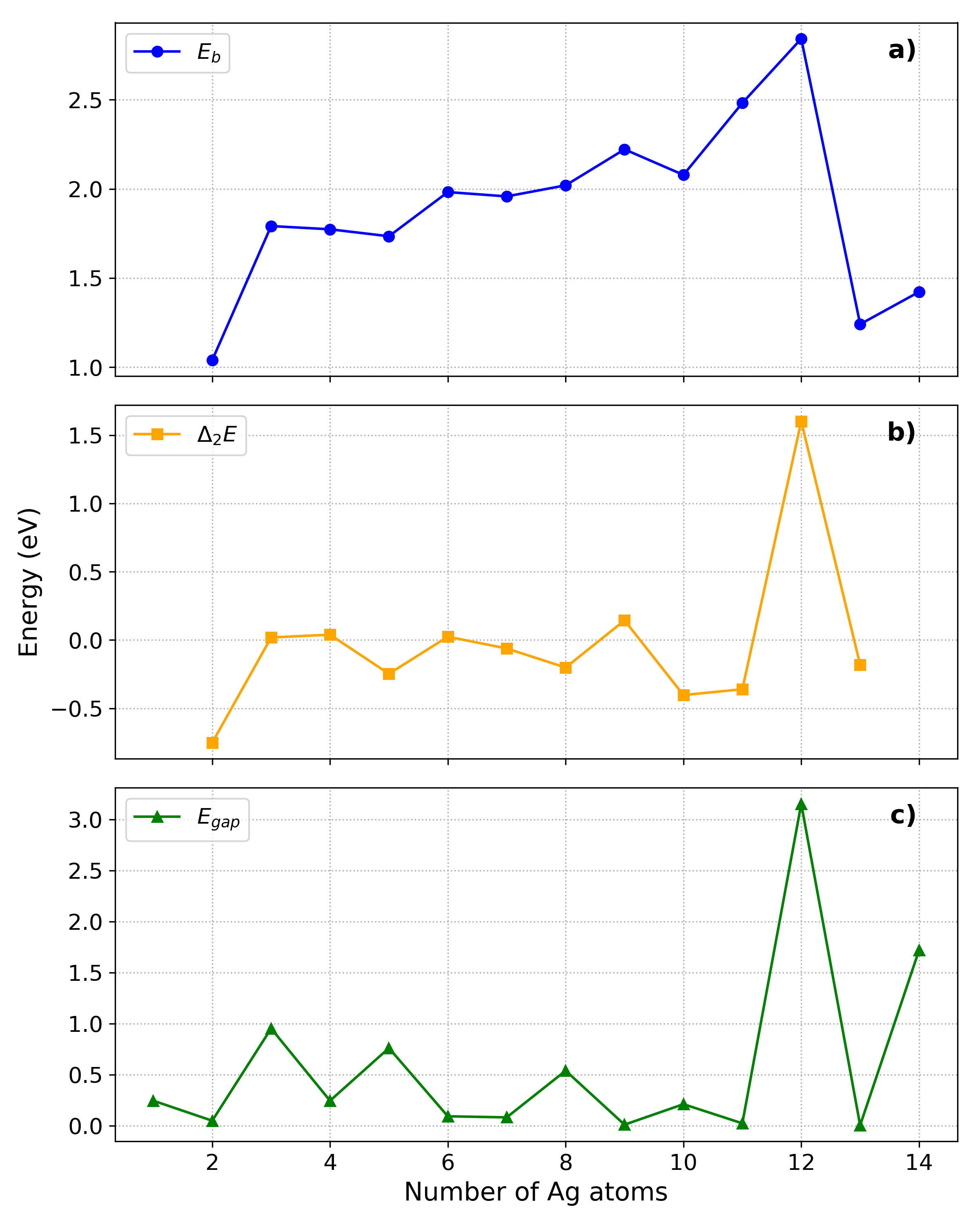}
    \caption{Binding energy ($E_b$), second-order energy difference ($\Delta_2 E$), and HOMO--LUMO gap ($E_{\text{gap}}$) as a function of cluster size for Ag$_n$Mo clusters.}
    \label{fig:AgnMo_properties}
\end{figure}

Figure~\ref{fig:AgnMo_properties} presents the binding energy ($E_b$), second-order energy difference ($\Delta_2 E$), and HOMO--LUMO gap ($E_{\text{gap}}$) as a function of cluster size for a series of Ag$_n$Mo clusters. The binding energy increases with cluster size, indicating enhanced stability as more Ag atoms are added, with a noticeable maximum at $n = 12$, suggesting a particularly stable configuration. The second-order energy difference $\Delta_2 E$ provides further insight into the relative stability between neighboring cluster sizes; the sharp peak at $n = 12$ confirms this size as a magic number, likely associated with a closed-shell or symmetric geometry. Interestingly, for small clusters, some $\Delta_2 E$ values are negative, which may reflect geometric rearrangements or electronic instabilities. The HOMO--LUMO gap remains generally small across most cluster sizes, consistent with metallic or semi-metallic behavior, but displays a pronounced peak at $n = 12$, further supporting enhanced electronic stability at this size. The correlation between high $E_b$, large $\Delta_2 E$, and wide $E_{\text{gap}}$ at $n = 12$ suggests that Ag$_{12}$Mo is a particularly stable and electronically distinct species in this series. These results align with previous stability analyses of metal-doped silver clusters.\cite{Tian2025}

\begin{figure}[htbp]
    \centering
    \includegraphics[width=0.47\textwidth]{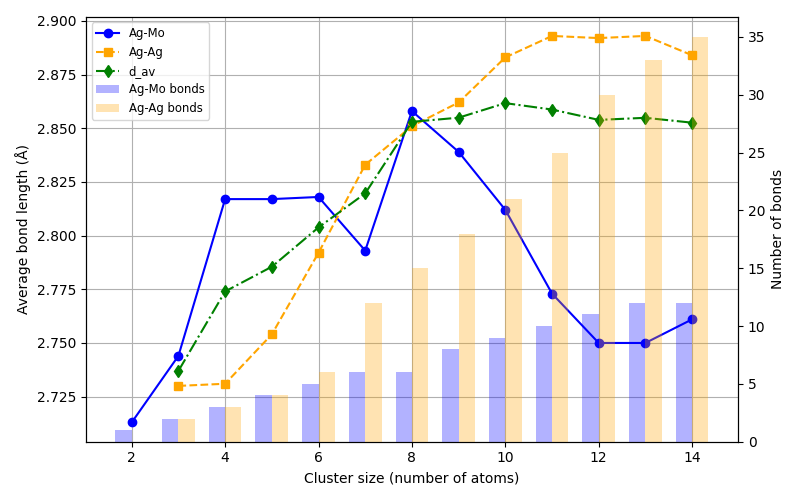}
    \caption{Average bond lengths and bond counts in Ag$_n$Mo clusters as a function of cluster size ($n = 2$--14). The curves show the average bond lengths for Ag--Mo (blue circles, solid line), Ag--Ag (orange squares, dashed line), and the overall weighted average bond length $d_{\mathrm{av}}$ (green diamonds, dash-dotted line). The bars (right axis) indicate the number of Ag--Mo (blue) and Ag--Ag (orange) bonds per cluster. Ag--Ag data starts from $n=3$, where at least two silver atoms are present.}
    \label{fig2}
\end{figure}

Figure~\ref{fig2} presents the evolution of average bond lengths and bond counts in MoAg$_n$ clusters as a function of cluster size ($n = 2$–14). The Ag--Mo bond lengths (blue circles) remain relatively stable across the series, with a slight increase up to $n=8$, followed by a mild contraction for larger clusters. In contrast, the Ag--Ag bond lengths (orange squares) exhibit a gradual and consistent increase with size, reflecting the progressive development of metallic bonding among Ag atoms as the clusters grow. The computed average bond distance $d_\mathrm{av}$ (green diamonds), defined as the bond-length-weighted average of all Ag--Mo and Ag--Ag interactions, provides a global descriptor of bonding within each cluster and interpolates smoothly between the two types of bonds. In parallel, the bars in the figure (referencing the right axis) indicate the number of bonds per cluster. The number of Ag--Mo bonds increases nearly linearly with size up to $n = 12$ and then saturates, while Ag--Ag bonds grow more sharply beyond $n=6$, indicating the densification and internal rearrangement of the Ag subnetwork. This structural transition highlights a shift from Mo-centered coordination toward more Ag-rich architectures, with Ag--Ag bonding becoming dominant in larger clusters. Altogether, the trends underscore the evolving interplay between heteronuclear (Ag--Mo) and homonuclear (Ag--Ag) interactions that shape the structural motifs as cluster size increases.

\begin{figure}[htbp]
    \centering
    \includegraphics[width=0.47\textwidth]{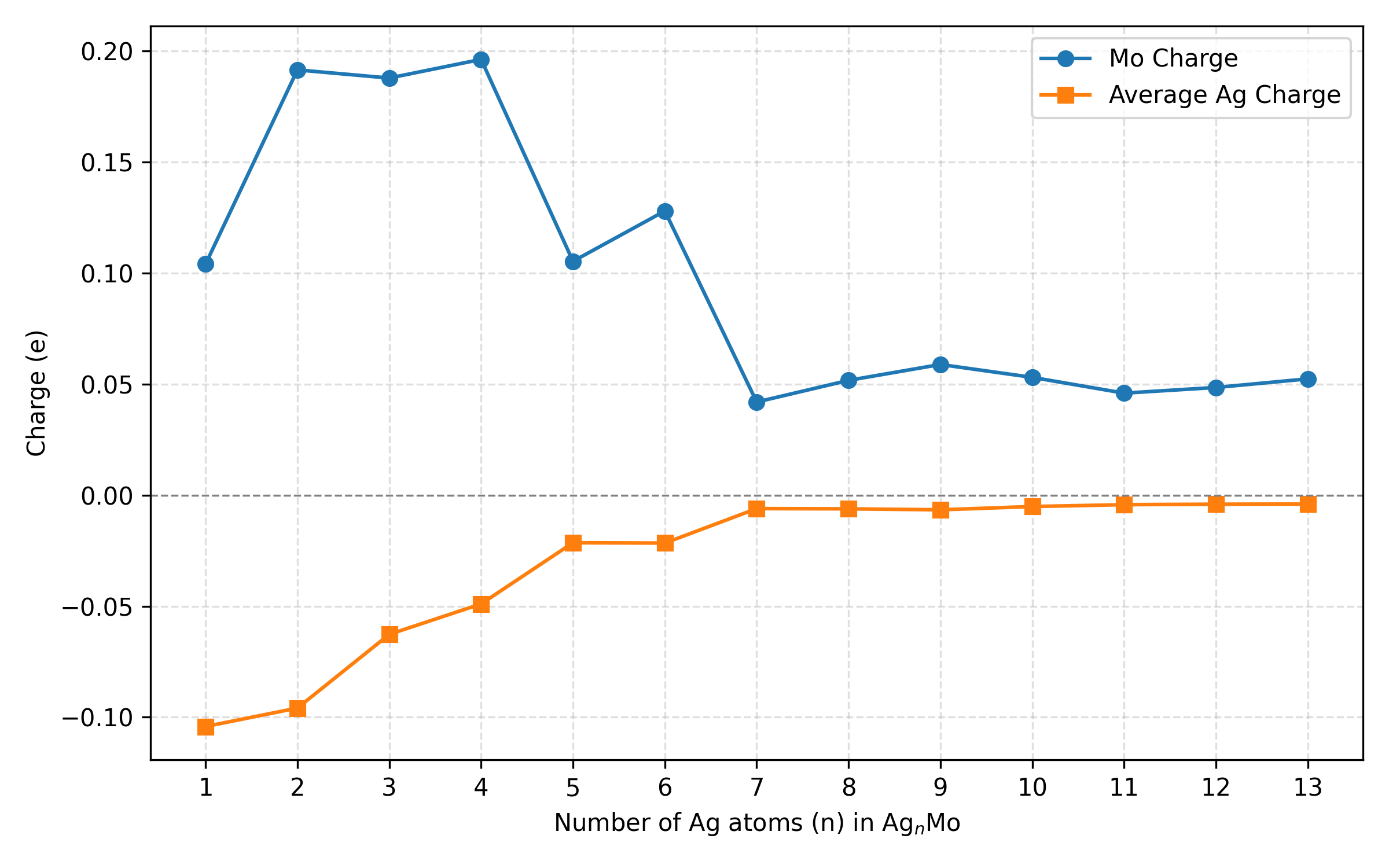}
    \caption{
        Hirshfeld charges on Mo (blue circles) and average Ag atom charges (orange squares) in Ag$_n$Mo clusters as a function of silver content ($n = 1$–14).    
    }
    \label{fig:hirshfeld_charge}
\end{figure}

The distribution of atomic charges in Ag$_n$Mo clusters ($n = 1$--14) was analyzed using the Hirshfeld scheme. Figure~\ref{fig:hirshfeld_charge} shows the computed net charge on the Mo atom (blue circles) and the average charge on Ag atoms (orange squares) as a function of silver content $n$. The total charge of each cluster is zero; thus, the partial charges reflect internal electron redistribution arising from bonding interactions. The Mo atom consistently bears a positive charge throughout the series, with values ranging from approximately 0.20~$e$ in small clusters ($n=3$--5) down to about 0.05~$e$ for $n\geq10$. This trend indicates a progressive decrease in net electron donation from Ag to Mo as the cluster grows, suggesting that in smaller clusters, the Mo center acts as a strong electron acceptor due to its higher electronegativity and central coordination. In larger clusters, the electron density becomes more delocalized, and the extent of charge localization on Mo is reduced. Conversely, the average charge per Ag atom is negative in all clusters, as expected due to the net charge conservation. However, its magnitude drops significantly with increasing $n$. For $n=1$, the Ag atom carries a charge of about $-0.10~e$, but this value decreases rapidly and stabilizes near $-0.004~e$ for $n\geq12$, indicating a transition toward a more metallic and less polarized bonding environment within the Ag subnetwork. Overall, this charge redistribution supports the structural evolution discussed previously: small Ag$_n$Mo clusters are characterized by localized Ag$\rightarrow$Mo charge transfer, while larger clusters exhibit more metallic character with delocalized bonding. The persistent, albeit reduced, positive charge on Mo suggests that it maintains a central coordination role even as the nature of bonding evolves with cluster size.

\begin{figure}[h!]
    \centering
    \includegraphics[width=0.47\textwidth]{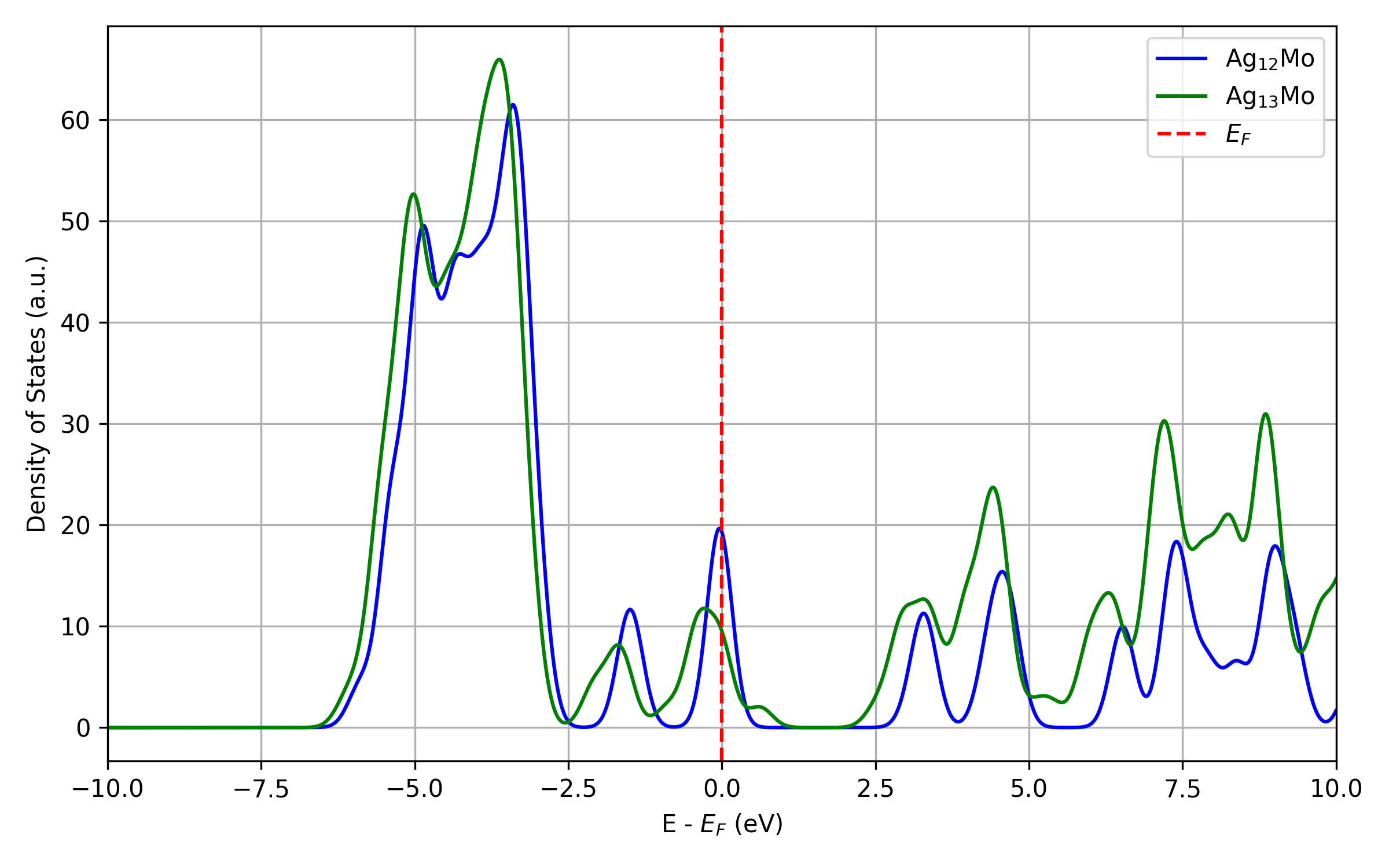}
    \caption{Density of States (DOS) for the clusters Ag$_{12}$Mo and Ag$_{13}$Mo, energies shifted such that the HOMO is at 0 eV (E$_F$ line indicated).}
    \label{fig:dos_comparison}
\end{figure}

Figure~\ref{fig:dos_comparison} compares the electronic density of states (DOS) for Ag$_{12}$Mo and Ag$_{13}$Mo clusters, with energies aligned to the HOMO level (set at 0 eV). Both clusters display prominent DOS features near the Fermi level, indicating accessible electronic states that may contribute to reactivity or conductivity. The DOS profile of Ag$_{12}$Mo is characterized by sharper and more discrete peaks, particularly below the Fermi level, suggesting a higher degree of electronic localization and possible shell closure effects, consistent with the icosahedral symmetry and enhanced stability previously observed for this cluster. In contrast, Ag$_{13}$Mo exhibits broader and more delocalized features, indicative of increased structural distortion and electronic hybridization due to the addition of a 13th Ag atom, which breaks the symmetry observed in Ag$_{12}$Mo. Notably, the sharper DOS peak just below the Fermi level in Ag$_{12}$Mo, absent or less pronounced in Ag$_{13}$Mo, suggests a potential difference in their electronic reactivity. These differences in DOS features correlate with the structural and energetic trends observed earlier, reinforcing the interpretation that Ag$_{12}$Mo represents a particularly stable and electronically distinct cluster in this size range.

\section{Conclusions}
In summary, our DFT investigations of Ag$_n$Mo clusters ($n=1$–14) elucidate the intricate interplay between structure, bonding, and electronic properties as a function of cluster size. We observe a clear structural evolution from planar, low-symmetry isomers in small clusters to compact, highly coordinated three-dimensional geometries in larger ones, with the icosahedral Ag$_{12}$Mo cluster standing out as a particularly stable and symmetric motif. The pronounced maxima in binding energy, second-order energy differences, and HOMO–LUMO gap at $n=12$ highlight this size as a magic number with enhanced stability and electronic shell closure. Bond length trends reveal that while Ag–Mo distances remain fairly constant, Ag–Ag bonds lengthen with size, consistent with increasing metallic character. Charge analysis confirms significant Ag-to-Mo electron transfer in small clusters, diminishing as cluster size grows and metallic bonding dominates. Collectively, these findings provide comprehensive insights into the size-dependent properties of Mo-doped silver clusters, contributing to the fundamental understanding necessary for their potential catalytic and nanomaterial applications.


\section{Acknowledgments}
P.L.R.-K. would like to thank the support of CIMAT Supercomputing Laboratories of Guanajuato and Puerto Interior.




\bibliographystyle{unsrt}
\bibliography{mendelei.bib}
\end{document}